\documentclass[aps,prl,twocolumn,groupedaddress,showkeys,showpacs]{revtex4}

\usepackage[dvips]{graphicx}

\begin{document}

\title{Granular segregation as a critical phenomenon}

\author{Pedro M. Reis}
\author{Tom Mullin}
\affiliation{Manchester Center for Nonlinear Dynamics,\\
University of Manchester, Oxford Road, Manchester, M13 9PL, UK}

\date{\today}

\begin{abstract}
We present the results of an experimental study of patterned
segregation in a horizontally shaken shallow layer of a binary
mixture of dry particles. An order parameter for the segregated
structures is defined and the effect of the variation of the
combined filling fraction, $\mathcal{C}$, of the mixture on the
observed pattern formation is systematically studied. The
surprising result is that there is a critical event associated
with the onset of the pattern, at $\mathcal{C}=0.647\pm0.049$,
which has the characteristics of a second order phase transition,
including critical slowing down.
\end{abstract}

\pacs{01.50Pa, 45.70Mg, 64.75.+g, 68.35.Rh, 45.70Qj.}

\keywords{Experiment, granular, binary mixture, segregation,
critical phenomena, pattern formation.}

\maketitle


Segregation is a counter-intuitive phenomenon where an initially
mixed state of dry granular particles separates into its
constituent components under excitation\cite{shinbrot:2001}.
Intriguingly, it does not always happen and the conditions for its
occurrence are difficult to predict. Segregation is not only of
fundamental interest but it is also of practical importance with
applications in areas ranging from industry\cite{bridgewater:1976}
to geology\cite{iverson:1997}. In recent years, there has been an
upsurge of interest in small scale laboratory studies where
vibration \cite{burtally:2002,rosato:1987}, avalanching in
partially filled horizontal rotating drums
\cite{gray:2001,choo:1997} and stratification in vertically poured
mixtures \cite{makse:1997} have all provided interesting examples
of pattern formation.

Our focus is on quasi-2-dimensional horizontally driven layers of
binary mixtures of particles \cite{mullin:2000,scherer:1996} as
this gives the practical advantage that any collective behavior is
readily visualized and gravity is effectively eliminated.
Moreover, the material is in contact with the drive throughout the
motion. For this class of binary granular systems a qualitative
segregation mechanism has been suggested \cite{aumaitre:2001}
borrowing the idea of \emph{excluded volume depletion} from
colloidal systems and binary alloys \cite{hill:1994}. In a driven
2D binary system of particles of different sizes, the packing
fraction of the system is decreased if self-organized clustering
of the larger particles occurs. Thereby considerably increasing
the number of configurational states of the system. Hence ordered
arrangement of the large spheres can increase the total entropy of
the system by increasing the mobility of the small particles. This
process has been referred to as \emph{entropic ordering}
\cite{shinbrot:2001}. We believe that analogous attractive
depletion forces are responsible for segregation in our granular
system. We show that segregation undergoes a phase transition and
occurs only for filling fractions above a critical value with the
characteristics of a second order phase transition with critical
slowing down.

          \begin{figure}[b]
                 \includegraphics[width=8.5cm]{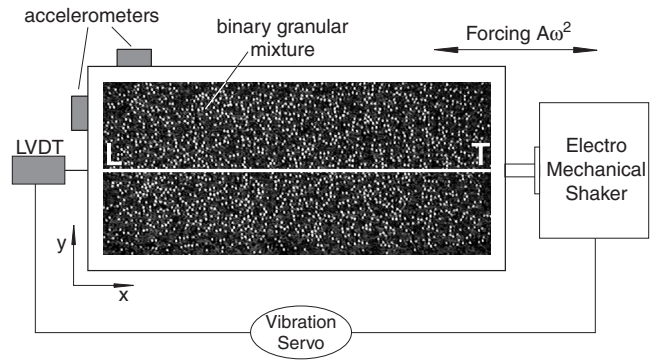}\\
\caption{\label{apparatus} Top diagrammatic view of the apparatus.
The experiment is set horizontally with gravity pointing into the
page. The inset photograph shows an example of the homogeneously
mixed initial conditions of the granular layer with
phosphor-bronze spheres (White regions = 1) and poppy seeds (Black
regions = 0).}

             \end{figure}

                 \begin{figure*}[t]
                    \includegraphics[width=15cm]{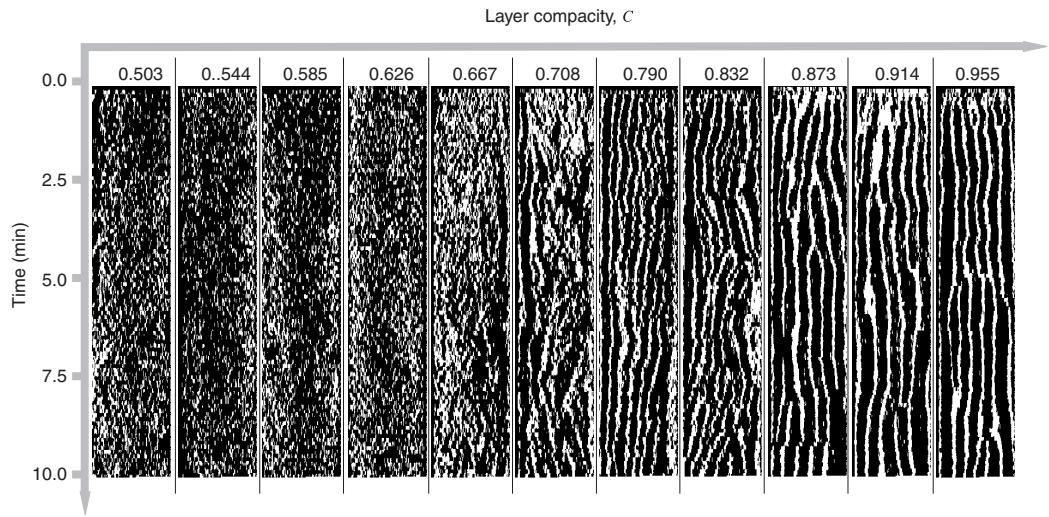}
\caption{\label{segregationdiagram} Series of space-time diagrams
for increasing $\mathcal{C}$. Each diagram is a stack, in time, of
the mi-layer cut line (LT in Fig. \ref{apparatus}). The LH side of
the diagram corresponds to \emph{gaseous} binary state. For larger
values of $\mathcal{C}$, on the RH side, segregated structures
form within the first minute of forcing. All runs were started
from homogeneously mixed initial conditions.}
                \end{figure*}


A schematic diagram of the top view of the apparatus is presented
in Fig. \ref{apparatus}. It consisted of a horizontal smooth
rectangular tray, of dimensions $(x,y)=180\times90mm$ with a
flatness of less than $\pm5\mu m$, on which a binary mixture of
particles was vibrated longitudinally. The tray, made out of
aluminum tool plate, was mounted on a system of four high
precision linear bearings and connected to a Ling
electro-mechanical shaker with a servo feedback control.   The
dynamical displacement and acceleration were monitored by a
linearly variable differential transducer (LVDT) and two
orthogonal piezoelectric accelerometers. The motion was checked to
be unidirectional and sinusoidal to better than 0.1\%. The forcing
parameters were kept constant at amplitude $A=\pm(1.74\pm0.01)mm$
and frequency of oscillation $\omega=12Hz$. Static charging
effects were eliminated by coating the container's surface with a
layer of colloidal graphite therefore making it conducting.
Furthermore, provision was made to evacuate the apparatus and this
had no influence on the segregated structures.

The binary mixture consisted of  poppy seeds
($\rho_{ps}=0.20gcm^{-3}$), which were approximately ellipsoidal
with major and minor axes $d^+_{ps}=1.07mm$ and
$d^{1-}_{ps}\approx d^{2-}_{ps}\approx0.50mm$, respectively, and
precision phosphor-bronze spheres ($\rho_{ps}=2.18gcm^{-3}$) with
diameter $d_{pb}=1.50mm$. The amounts used  were always such that
a shallow layer regime was maintained, with layer height
$h\leq1.5mm$.

Forcing of individual particles arose principally from the
frictional interaction with the base of the container and the
effects of boundary walls were observed to be localized. Although
the motion of the container was sinusoidal, the stick-and-slip
nature of tray/particle contact induced stochastic thermalization;
quasi-2-dimensional trajectories were obtained in $x$ and $y$,
with the characteristics of a 2D random walk \cite{reis:2002}. The
surface of the tray was accurately levelled with fine-pitch
adjustable screws to within $\pm0.0086^\circ$ and checked using a
high precision engineers level. This ensured homogeneous forcing
across the whole layer.

The granular layer was imaged using a CCD camera mounted directly
above the apparatus and $600\times300$ pixels frames were
digitized into a PC. Digital image processing techniques were
employed to transform the acquired video frames into binary
format, giving a value of 0 (black) to pixels corresponding to
regions of poppy seeds and a value of 1 (white) for those of
phosphor-bronze spheres. Frames were thereby analyzed as
$600\times300$ digital matrices, allowing measurements on the
dynamics of the granular patterns to be performed.

We define a measure of the combined filling fraction of the layer,
which we call the \emph{layer compacity}, to be,
\begin{equation}
\mathcal{C}=\frac{N_{ps}A_{ps}+N_{pb}A_{pb}}{xy},
\end{equation}
where $N_{ps}$ and $N_{pb}$ are the numbers of poppy seeds and
phosphor-bronze spheres in the monolayer, $A_{ps}=0.87\pm
0.09mm^2$ and $A_{pb}=1.77\pm 0.06mm^2$ are the two dimensional
projected areas of the respective individual particles and $x$ and
$y$ are the longitudinal and transverse dimensions of the tray. We
regard $\mathcal{C}$ as the principle parameter of the system.

A digitized image of a mixture of the two types of particles  has
been superposed on the schematic diagram of the apparatus in Fig.
\ref{apparatus}. It is representative of the initial conditions
which were consistently set using the following method. Firstly,
$N_{ps}$ poppy seeds were vibrated at large amplitudes, $A\sim\pm
5mm$, creating an homogeneous and isotropic sub-monolayer. The
phosphor-bronze spheres were then suspended, at a height $h=15mm$
above the poppy seeds layer, on a horizontal frame with a
perforated array of ($57\times28$) holes, of $2mm$ diameter,
arranged in a triangular lattice with a cell spacing of
$a=2.75mm$. A shutter was then opened and the 1596 phosphor-bronze
spheres fell onto the layer of poppy seeds, creating a near
homogeneous mixture of the two types of particles.


The focus of the investigation was to systematically increase
$\mathcal{C}$, consequently decreasing the mobility of individual
particles, and explore the conditions under which granular
segregation took place. To achieve this, we incrementally
increased the number of poppy seeds, $N_{ps}$, in the layer, in
measured steps, with the amount of phosphor-bronze spheres,
$N_{pb}$, held constant. We also experimented with changing the
number of phosphor-bronze spheres and found that this did not
qualitatively affect the main findings and the quantitative
changes were found to be linearly scalable. We chose to perform
the experiments by changing the numbers of poppy seeds since this
provided finer steps in the control parameter, i.e.
$\mathcal{C}=\mathcal{C}(N_{ps})$.

In Fig.~\ref{segregationdiagram} we present a series of space-time
diagrams, for increasing $\mathcal{C}$, which were constructed by
sampling  the mid-frame cut line along the granular layer's
$x$-dimension (shown in Fig. \ref{apparatus} as the white
longitudinal line $LT$), and progressively stacking, in time, 200
of these ($600\times1$ pixels) lines (acquisition rate of
$1/6Hz$), for runs of 10min. This period was found to be long
enough for steady state conditions to be achieved.

          \begin{figure}[t]
                \includegraphics[width=5.5cm]{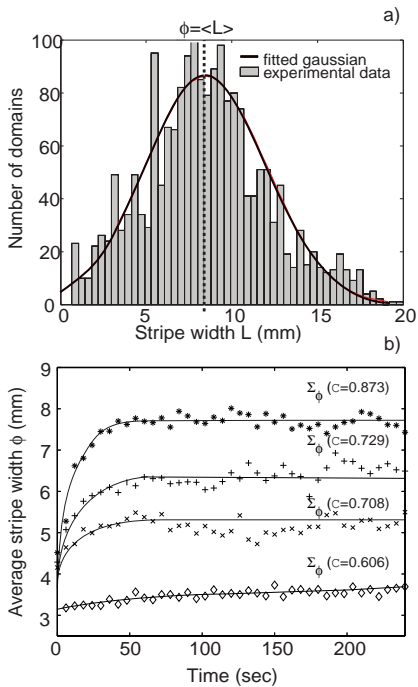}
\caption{\label{phi} a) Histogram of the $L$ distribution
(longitudinal stripe widths) for a single $600\times300$ pixel
frame ($\mathcal{C}=0.966$). Solid curve is Gaussian fit. Dashed
line is average value $\phi=\langle L\rangle$ which we define to
be the order parameter. b) Time series for the order parameter
$\phi$ for various values of $\mathcal{C}$. A saturation level is
attained after a fast initial segregation period. Solid curves are
fits to Eqn. (\ref{fits}). }
             \end{figure}

In the left-hand image of Fig.~\ref{segregationdiagram}, with
$\mathcal{C}=0.503\pm0.035$, it can be seen that the  binary layer
remains mixed for the duration of the experiment. The system acts
as a low density granular gas with two components and individual
particles describe quasi-2-dimensional Brownian-like trajectories.
Similar behavior has been found in numerical studies of
two-dimensional binary granular gases with uniform thermalization
in which a mixed non-equilibrium steady state is achieved at low
filling fractions \cite{henrique:2000}. The right-hand image in
Fig.~\ref{segregationdiagram}, for $\mathcal{C}=0.955\pm0.081$,
displays qualitatively different behavior. It can be seen that
within the first minute of the start of the motion, small
localized clusters of phosphor-bronze spheres (white regions) form
and progressively coalesce with neighbouring ones. Thereby
coarsening occurs, so that well defined stripes are eventually
formed, aligned perpendicularly to the direction of forcing
\cite{mullin:2000}.

At intermediate values of $\mathcal{C}$, partially segregated
states emerge and we aim to show that there is a critical
dependence of the process on the layer compacity rather than a
smooth emergence of the segregated structures. Thus there is a
critical value, $\mathcal{C}_c$, below which the binary granular
layer remains mixed. Above $\mathcal{C}_c$, as the layer is
incrementally compacted, segregation structures develop in a
nonlinear manner. Specifically, we treat the process as if it were
a second order phase transition.



In order to proceed, we first establish an order parameter,
$\phi$, which we choose to be the characteristic longitudinal
width of the phosphor-bronze (white) stripes. For each acquired
frame this was calculated by scanning through each of the 300
horizontal lines; the length of the \emph{white steps} was
measured, $L(^{0\rightarrow 1}_{1\rightarrow 0})$, which typically
yielded a distribution of $\sim6000$ identified stripe widths.
These were distributed normally, as shown in Fig. \ref{phi}a,
where a fitted Gaussian function has been superimposed on the
experimental histogram of $L$. Since the distribution is well
defined, we have chosen the order parameter to be $\phi=\langle L
\rangle$, i.e. the average value for the distribution of
longitudinal widths of the segregated stripes.

\begin{figure}[b]
 \includegraphics[width=7cm]{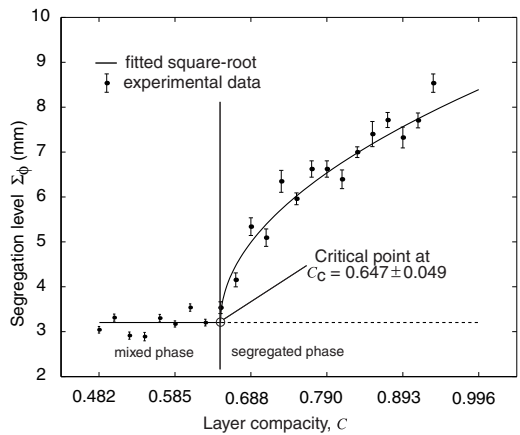}
  \caption{\label{sqrtsingularity}  Phase transition plot for segregation level. Solid line is theoretical curve for square-root singularity.
  Critical point occurs at $\mathcal{C}_c=0.647\pm0.049$. The mixed phase exists below this point and the segregated phase above it.}
  \end{figure}


In Fig. \ref{phi}b we present time series of $\phi$ for various
$\mathcal{C}$, where each run has been started from mixed initial
conditions. At large compacities, e.g.
$\mathcal{C}=0.914\pm0.077$, there is a fast initial growth,
eventually saturating after $\sim 1min$ from the start of the
experiment. At this point an approximately constant level is
attained. The superimposed solid lines are fits of the form,
        \begin{equation}
            \phi^{fit}(\mathcal{C},t)=\Sigma_\phi(\mathcal{C}) -
            \alpha.\exp\left(-\frac{t}{t_s(\mathcal{C})}\right),
            \label{fits}
        \end{equation}
to the experimental time series of $\phi$ where
$\Sigma_\phi(\mathcal{C})$, the value at which $\phi$ saturates,
is the \emph{segregation level}, $\Sigma_\phi(\mathcal{C})-\alpha$
is the initial value, $\phi(t=0)$, and $t_s(\mathcal{C})$, the
time scale associated with the saturation of the order parameter,
is the \emph{segregation time} .

The $\mathcal{C}$-dependence of the segregation level is presented
in Fig. \ref{sqrtsingularity}.  We see that for low compacities,
up to a value, $\mathcal{C}_c=0.647\pm0.049$, the segregation
level remains constant at $\langle \Sigma_\phi \rangle =
3.21\pm0.08 mm$. This corresponds to approximately  two sphere
diameters so that any domains are not classified as structures and
the system is deemed to be in a mixed or unsegregated state. As
$\mathcal{C}$ is increased past $\mathcal{C}_c$, segregated
stripes of increasing $\phi$ emerge and the segregation level
exhibits a square-root singularity. The solid curve in Fig.
\ref{sqrtsingularity} is the line of best fit of the form,
            \begin{equation}
                \Sigma_\phi^{fit}=A(\mathcal{C}-\mathcal{C}_c)^{\beta}
                +\Sigma_\phi(\mathcal{C}_c),
            \end{equation}
where $A$ is the transition scaling factor,
$\{\mathcal{C}_c,\Sigma_\phi(\mathcal{C}_c)\}$ is the critical
point at which the transition occurs and $\beta=\frac{1}{2}$ is
the order parameter exponent. The numerical values for the fitted
parameters were $A=2.81\pm0.07mm$, $\mathcal{C}_c=0.647\pm 0.049$
and $\Sigma_\phi(\mathcal{C}_c)=3.21\pm0.08mm$.

              \begin{figure}[t]
                 \includegraphics[width=7cm]{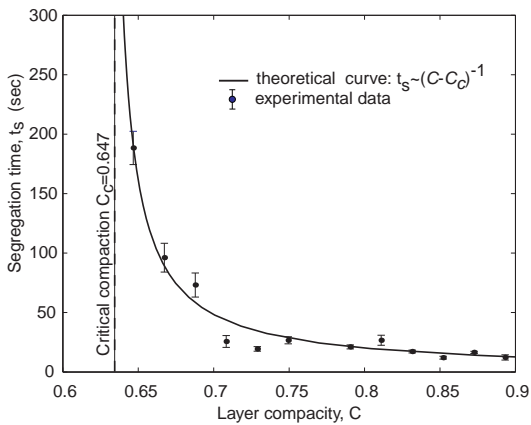}\\
                 \caption{\label{criticalts} Plot of segregation time, $t_s$, which diverges as the
                 critical point $\mathcal{C}_c$ is approached from above. Solid line is theoretical curve of the
                 form: $t_s\sim(\mathcal{C}-\mathcal{C}_c)^{-1}$.}
             \end{figure}

It is interesting  to note that the generally accepted value of
maximum packing fraction for a random closed packing for a 2D
system of particles  \cite{rouille:1990} is around 0.82 which is
considerably larger than  $\mathcal{C}_c$ for segregation in our
system. This seems to indicate that the segregation transition
occurs while the granular mixture is still in the monolayer
regime.


A further  measure that supports  the notion of critical behavior
is the $\mathcal{C}$-dependence of the segregation time, $t_s$,
extracted from the fittings of $\phi(\mathcal{C},t)$  to the
function in Eqn. (\ref{fits}). Note that $t_s$ is only defined for
$\mathcal{C}>\mathcal{C}_c$, corresponding the region where
segregation occurs. The measured segregation time is plotted in
Fig. \ref{criticalts} as a function $\mathcal{C}$. As expected for
a second order transition, $t_s$ diverges, as $\mathcal{C}$ is
decreased from above, near $\mathcal{C}_c$. This is usually
referred to as \emph{critical slowing down} where the divergence
has the form of $t_s\sim(\mathcal{C}-\mathcal{C}_c)^{-\gamma}$
\cite{chaikin:1995}. The solid line in Fig. \ref{criticalts}
corresponds to the mean field approximation result,
$\gamma=1$~\cite{anisimov:1991}.


In summary, we have shown that granular segregation in a
horizontally shaken layer displays both qualitative and
quantitative features which are consistent with a second order
phase transition, i.e. there is a critical layer compacity value
below which mixing of the binary layer is guaranteed and above
which reproducible segregation patterns form.  The phenomena were
not affected when the granular layer was evacuated, establishing
that this is a fundamental process intrinsic to the dynamics of
the granular layer. The critical exponents for the order parameter
and the segregation time agree with those of mean field type
behavior, $\beta=\frac{1}{2}$ and $\gamma=1$, respectively. This
is surprising since the system is far from equilibrium; it is both
forced and the principal energy losses are due to inter-particle
inelastic collisions. A reconciliation between this results on
granular segregation and recent theoretical reports of critical
behavior in granular hydrodynamics \cite{santos:2001} and the
existence of self-organized domains in granular compaction models
\cite{baldassarri:2001} may be crucial to our further
understanding of segregation and other collective phenomena in
granular materials.

\bibliography{paper}

\end{document}